\documentclass[pre,twocolumn,showpacs,preprintnumbers,amsmath,amssymb,superscriptaddress]{revtex4-1}

\usepackage{graphicx}
\usepackage{dcolumn}
\usepackage{bm}
\usepackage{epsfig}

\begin{document}


\title{Fragile to strong crossover coupled to liquid-liquid transition in hydrophobic solutions}

\author{D. Corradini}
\affiliation{Center for Polymer Studies and Department of Physics,
  Boston University, 590 Commonwealth Avenue, Boston, Massachusetts
  02215, USA} 
\author{P. Gallo}
\email{gallop@fis.uniroma3.it}
\affiliation{Dipartimento di Fisica, Universit\`{a} Roma Tre, 
Via della Vasca Navale 84, I-00146 Roma, Italy}
\author{S. V. Buldyrev}
\affiliation{Department of Physics, Yeshiva University, 500 West 185th
  Street, New York, New York 10033, USA} 
\author{H. E. Stanley}
\affiliation{Center for Polymer Studies and Department of Physics,
  Boston University, 590 Commonwealth Avenue, Boston, Massachusetts
  02215, USA} 

\date{7 May 2012}

\begin{abstract}

\noindent
Using discrete molecular dynamics simulations we study the relation
between the thermodynamic and diffusive behaviors of a primitive model
of aqueous solutions of hydrophobic solutes consisting of hard spheres
in the Jagla particles solvent, close to the liquid-liquid critical
point of the solvent.  We find that the fragile-to-strong dynamic
transition in the diffusive behavior is always coupled to the
low-density/high-density liquid transition. Above the liquid-liquid
critical pressure, the diffusivity crossover occurs at the Widom line,
the line along which the thermodynamic response functions show
maxima. Below the liquid-liquid critical pressure, the diffusivity
crossover occurs when the limit of mechanical stability lines are
crossed, as indicated by the hysteresis observed when going from high to
low temperature and vice versa.  These findings show that the strong
connection between dynamics and thermodynamics found in bulk water
persists in hydrophobic solutions for concentrations from low to
moderate, indicating that experiments measuring the relaxation time in
aqueous solutions represent a viable route for solving the open
questions in the field of supercooled water.

\end{abstract}



\maketitle 

\section{Introduction}

Over the past decade our understanding of the puzzling behavior of water
in the supercooled region has greatly expanded
\cite{block1}.  
At present, plausible scenarios that might fill
in the missing pieces of our comprehension have been developed through
simulations and theoretical models.
In 1992, using a molecular dynamics simulation of the ST2 potential, the
existence of a liquid-liquid critical point (LLCP) in the supercooled
region of water was hypothesized \cite{poole}. Since
that time the LLCP has been found in a number 
of other simulations, see for example Refs.~\cite{block2,xu,corradini10jcp}.
In this conceptual framework the presence of the LLCP
explains the puzzling anomalies in the behavior of water.  Although the
LLCP scenario remains fascinating, its existence has not been
experimentally confirmed, and alternative scenarios have been proposed
\cite{block3}.
Experiments are consistent with the existence of a LLCP
\cite{block4,mishima} but proof has been difficult because
efforts to reach the zone in the phase diagram where the LLCP would be
located are hampered by nucleation. On the low
temperature side 
the existence of high density amorphous (HDA) and low density amorphous
(LDA) states with a first-order coexistence line separating them
has been experimentally proven \cite{block5}.
If a LLCP exists, it is at the terminating point of the extension of this line in the supercooled liquid region. 
\cite{mishima,block6}.

In 1996 a strong relationship between the dynamic and thermodynamic
behaviors of water was reported \cite{sciortinogallo96},
indicating that the well-known singular temperature $T_S$~\cite{speedy} at
which extrapolation of various thermodynamic and dynamic anomalies diverge, can be identified with
the crossover temperature $T_C$ of the mode coupling theory
\cite{gotze2010}.
This crossover temperature marks the fragile-to-strong (FTS)
transition that occurs in supercooled liquids approaching the glass
transition.  Recent studies have completed the picture for water by
connecting this FTS transition to the crossing of the
``Widom line'' \cite{xu,block7,WL2}.  Upon approaching
the LLCP, the lines of maxima of the different response functions converge
on this Widom line, which separates water with HDL-like features at high
temperatures from water with LDL-like features at low temperatures
\cite{WL1,WL2}.  

Simulations have indicated that
the LLCP persists in aqueous solutions for concentrations ranging from
low to moderate both for polar and apolar solutes
\cite{corradini10jcp,corradini08,corradini11,corradini10pre,chatterjee}. This suggests that
focusing on aqueous solutions might be a possible experimental strategy
in probing the existence of a LLCP. 
In solutions, in fact, the nucleation barrier can be tuned and brought
below the LLCP~\cite{corradini10jcp,corradini11}.
Very recently experiments on water/glycerol mixtures~\cite{murata} have shown 
signs of a liquid-liquid phase transition.
Nonetheless, thermodynamic measurements are more difficult than 
measurements on dynamics, therefore it is timely to prove whether the
FTS crossover remains coupled to 
the liquid-liquid transition also in aqueous solutions, as this
can address the experimental quest on the existence of LLCP
toward the study of dynamics. We note that it
is a priori not obvious that the FTS transition
transfers to aqueous solutions as it is a phenomenology
present in many glass formers that do not have a LLCP.
In this paper we use a primitive model able to reproduce the phenomenology of
water.  We present the results of discrete molecular dynamics (DMD)
simulations on water mixed with hard spheres (HS), in order to determine
whether the connection between dynamics and thermodynamics retains its
validity in hydrophobic solutions.

The paper is structured as follows. Simulation details are reported in
Sec.~\ref{sim}. The results are presented and discussed in Sec.~\ref{res}.
Concluding remarks are given in Sec.~\ref{conc}.

\section{Simulation details}\label{sim}

The primitive model used for the results presented in this paper is the
Jagla model \cite{jagla} [see Fig.~\ref{fig:1}(a)], an isotropic
potential with two length scales, a hard-core distance $a$ and a
soft-core distance $b$, plus an attractive ramp that extends to a cut-off distance $c$.  
The parameterization we use,
$b/a=1.72$, c/a=3 and $U_R/U_0=3.56$, has been shown to possess bulk
water anomalies and a liquid-liquid transition \cite{xu,xu2}.  
The Jagla model was also successful in reproducing the increase of
solubility upon cooling of non-polar solutes modeled as HS of diameter $a$~\cite{buldyrev07,maiti}.

Physical
quantities are expressed in reduced units \cite{corradini10pre}.
The total number of particles contained in the cubic
simulation box is $N=1728$.
We performed DMD simulations on the discretized version of the Jagla 
ramp potential, for
the bulk and for $x_{\rm HS}=0.10$ and $x_{\rm HS}=0.15$ HS(J) mixtures of HS in
Jagla particles, at constant $N$, $P$ and $T$.  Pressures and
temperatures are controlled by Berendsen algorithms. The
HS have diameter $a$ and the same mass as the Jagla
particles. Solute-solute and solute-solvent interactions are purely
hard-core.  The position of the LLCP of the bulk Jagla system
\cite{xu,xu2,xu3} and of HS(J) mixtures with mole fraction up to
$x_{\rm HS}=0.50$ has been previously determined \cite{corradini10pre} in
$NVT$ simulations.  Its position in the $P-T$ plane shifts to higher
pressures and lower temperatures upon increasing the mole fraction of
HS.  
In the Jagla model the slope of the liquid-liquid coexistence line is positive,
unlike other models for water~\cite{block1}. Consequently the HDL is 
the more ordered phase and has an Arrhenius dynamic behavior, while the reverse is true
for other models~\cite{xu,corradini10pre,xu2}.
Simulations with primitive models have the advantage of permitting
to explore for large systems, a wide range of pressures and temperatures, especially
in the deep supercooled region, where equilibration of long-range, orientational
dependent potentials becomes impossible.
In particular, the ramp potential with the choice of parameters used here is able to reproduce, 
{\it mutatis mutandis}, the complex thermodynamic scenario of water 
commonly derived from long-range orientational dependent potentials~\cite{poole,block2,corradini10jcp}.

\section{Results and Discussion}\label{res}
 
\begin{figure}[t] 
\begin{minipage}{\columnwidth}
\psfig{file=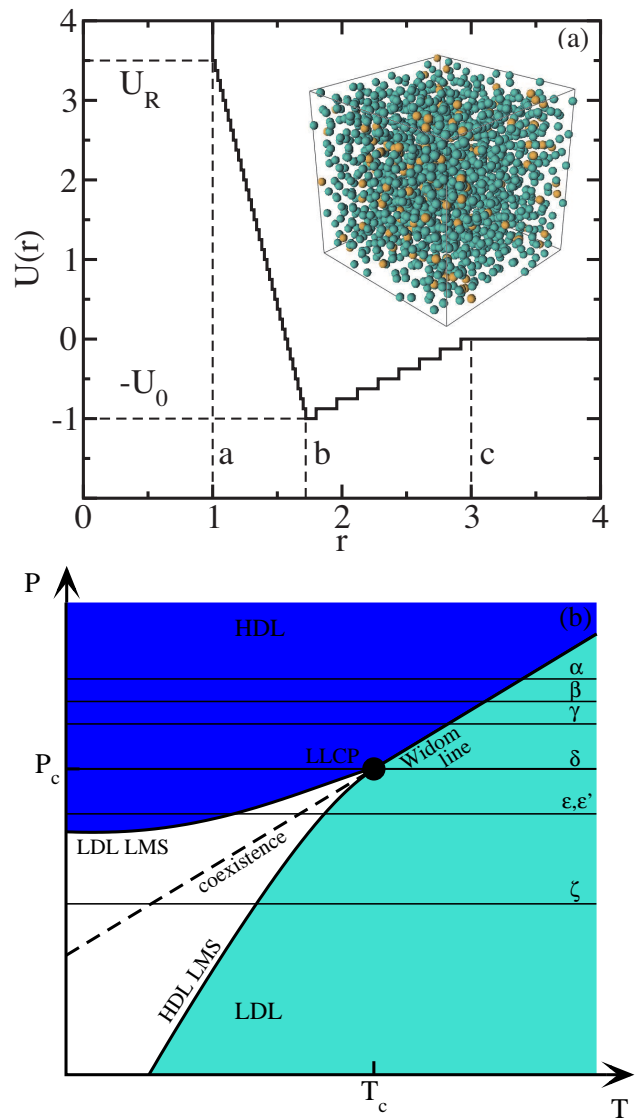,width=0.95\columnwidth}
\end{minipage}
\begin{minipage}{\columnwidth}
\hspace{0.4cm}
\psfig{file=fig1b.eps,width=0.94\columnwidth}
\end{minipage}
\caption{(Color online) (a) Spherically symmetric Jagla ramp potential. A snapshot of
  the $x_{\mbox {\scriptsize HS}}$ = 0.10 system is also shown in the
  top right corner. 
(b) Schematic liquid-liquid phase diagram for Jagla systems
  (bulk or solutions) with the LLCP and the limit of mechanical stability (LMS) 
  lines.  The constant pressure paths are simulated at
  $P=P_c+0.020$ (path $\alpha$); $P=P_c+0.015$ ($\beta$); $P=P_c+0.010$
  ($\gamma$); $P=P_c$ ($\delta$); $P=P_c-0.010$ ($\epsilon$) starting
  from high temperature; $P=P_c-0.010$ ($\epsilon '$) starting from low
  temperature; $P=P_c-0.030$ ($\zeta$).}
\label{fig:1}
\end{figure}

\begin{figure*}[!t] 
\centerline{\psfig{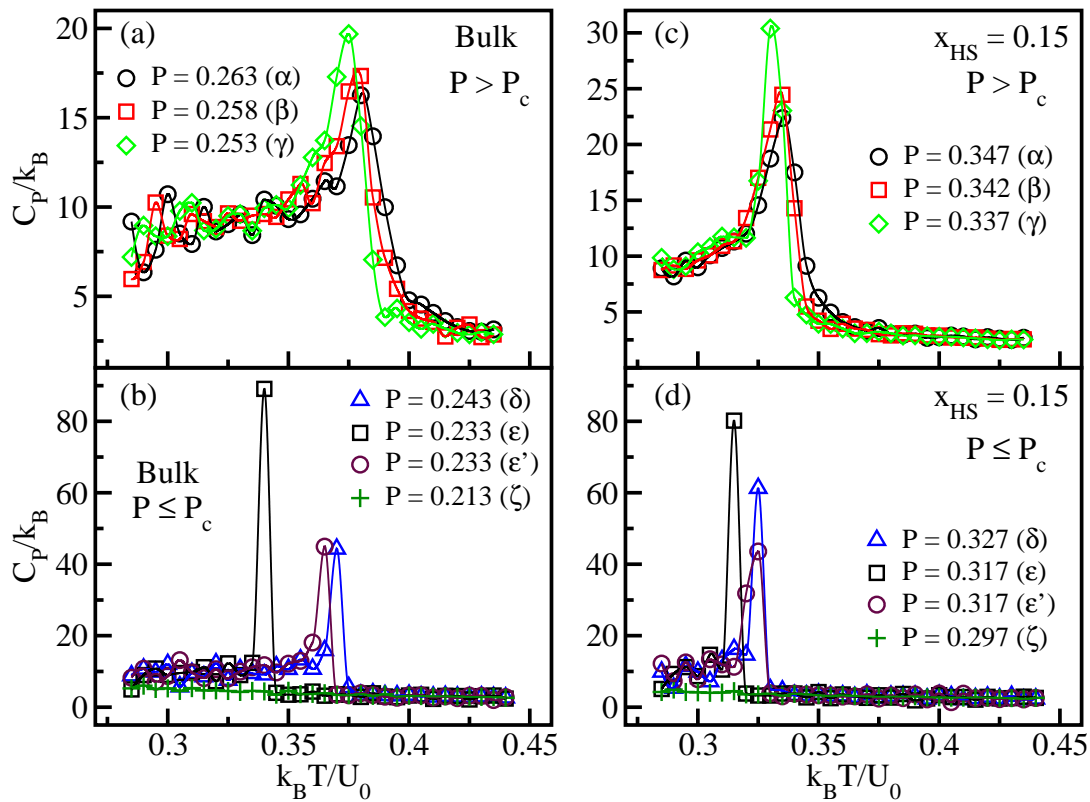}}
\caption{(Color online) Isobaric specific heat $C_P$ as a function of temperature (a,b)
  of bulk Jagla particles; (c,d)  of the $x_{\mbox {\scriptsize HS}}=0.15$ HS(J) solution.
  Top panels are for paths with $P>P_c$, bottom panels for paths with
  $P\le P_c$. Lines are cubic splines to the points.
  }  
\label{fig:2}
\end{figure*}

\begin{figure*}[!t] 
\centerline{\psfig{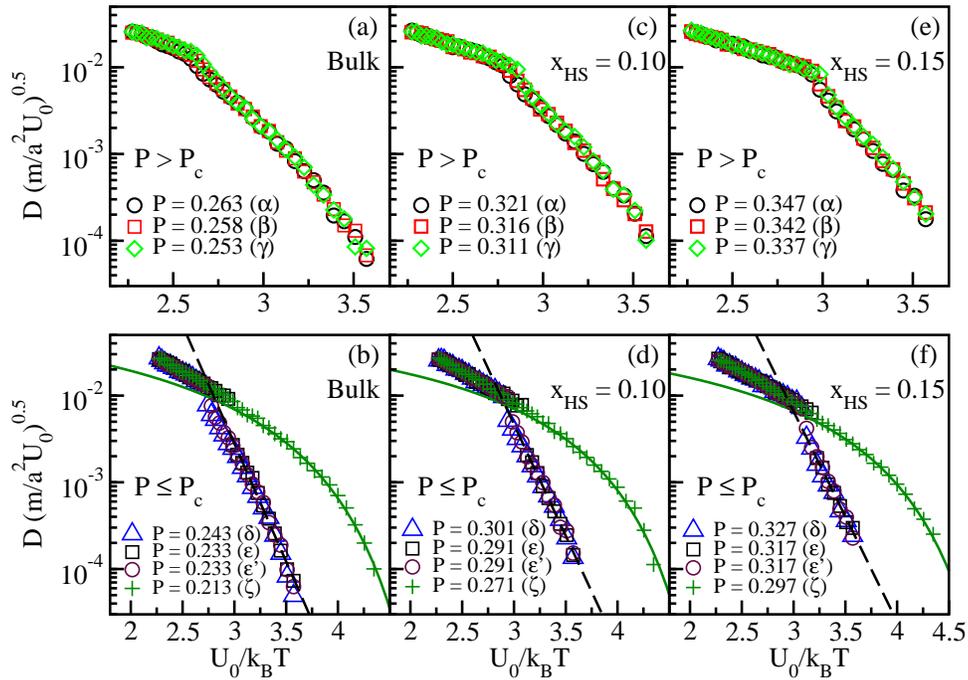}}
\caption{(Color online) Diffusion coefficient $D$ as a function of inverse temperature
  (a,b) of bulk Jagla particles; (c,d) of the $x_{\mbox {\scriptsize HS}}=0.10$ 
  HS(J) solution; (e,f) of the $x_{\mbox {\scriptsize HS}}=0.15$ HS(J) solution.
  Top panels are for paths with $P>P_c$, bottom panels are for paths with 
  $P\le P_c$. The solid curve is a VFT fit $D=D_0\exp[-B/(T-T_0)]$ for paths $\zeta$.
The dashed straight line is an example of Arrhenius fit $D=D_0\exp(-E_A/k_BT)$ for paths
  $\epsilon$. 
  The parameters of the fits are $D_0\simeq 0.4$, $B\simeq 0.25$, $T_0\simeq 0.18$ for VFT
   and $D_0\simeq 10^5$, $E_A\simeq 6$ for Arrhenius, for all systems. 
   }
\label{fig:3}
\end{figure*}

In order to verify the connection between dynamic and thermodynamic
behaviors, we consider the phase diagrams of bulk water and aqueous
solutions previously determined for the Jagla potential and we study the
thermodynamic and diffusive behavior along constant pressure paths [see
  Fig.~\ref{fig:1}(b)]. In particular, we consider isobaric paths near
and above the LLCP, at the LLCP, and near and below the LLCP.  Paths
$\alpha$, $\beta$, and $\gamma$ are above the critical pressure and
cross the Widom line. The path $\delta$ is at the critical
pressure. Paths $\epsilon$ and $\epsilon '$ are below the critical
pressure, and cross the HDL and LDL limit of mechanical stability (LMS)
lines depending on the direction. 
In particular, along path $\epsilon$ the system goes
from LDL (high $T$) to HDL (low $T$)  when crossing the LDL LMS while along path $\epsilon
'$ it goes from HDL to LDL upon crossing the HDL LMS.  Path $\zeta$, also below the critical pressure, does not cross
the LDL LMS line and therefore the system remains in the LDL
all along the path.  All paths, except path $\epsilon '$, have been
performed starting from a configuration at high temperature and
progressively equilibrating the system at lower temperatures. For path
$\epsilon '$ the starting configuration was taken at low temperature and
the system was progressively equilibrated at increasing temperatures in
order to cross also the HDL LMS.

Figure~\ref{fig:2} shows the behavior of the isobaric specific heat
$C_P\equiv (1/N)\cdot (\partial H/\partial T)_P$ for bulk water and for
the $x_{\rm HS}=0.15$ HS(J) solution calculated for all the paths shown in
Fig.~\ref{fig:1}(b).  The results for the $x_{\rm HS}=0.10$ HS(J) solution
follow a similar trend and are not shown.  Both in bulk water and in the
solutions, the points above the critical pressure at which the specific
heat displays a maximum define the Widom line emanating from the LLCP.
Consistent with the positive slope of the liquid-liquid coexistence line
for the Jagla potential \cite{xu,xu2}, the specific heat maximum moves
to a higher temperature as the pressure increases. 
Its height decreases moving away from the LLCP.  
For the paths along the critical pressure, 
the temperature of the maximum falls close to the one estimated in
$NVT$ simulations \cite{corradini10pre, xu2}. We note that paths $\epsilon$
and $\epsilon '$ clearly show the hysteresis expected for a first-order
transition in a finite size system with the $C_P$ maxima occurring at
different temperatures, one corresponding to the LDL LMS for path
$\epsilon$ and one corresponding to the HDL LMS for path $\epsilon '$.
Along path $\zeta$, which does not reach the LDL LMS, no peak is
observed.

We now show the diffusive behavior of Jagla particles in solution along
the same constant pressure paths and compare it with the bulk Jagla
results.  Figure~\ref{fig:3} shows the diffusion
coefficient as a function of inverse temperature for bulk Jagla
particles and $x_{\rm HS}=0.10$ and $0.15$ HS(J) solutions. 
For all
three systems, above the LLCP the diffusion coefficient shows a
crossover as it crosses the Widom line as determined from the $C_P$
maxima.  
Along path $\zeta$, that does not cross the LDL LMS line, the temperature dependence
of $D$ is clearly non-Arrhenius, indicating that the LDL liquid can be classified as fragile.
Taking advantage of this particular potential we could push this path to 
very low temperatures and thus observe a clear fragile behavior.
In Fig.~\ref{fig:3} we also show the fit to the Vogel-Fulcher-Tamman,
VFT, function $D=D_0\exp[-B/(T-T_0)]$ characterizing the fragile behavior of a glass former.

For all the paths crossing 
the Widom line, on the high temperature side of this line we find a
LDL-like liquid, and the diffusive behavior 
is that of a fragile
liquid, when compared to path $\zeta$ which is always in the LDL fragile environment.  On the low temperature side of the Widom
line the diffusive behavior shows a steep 
straight slope typical of a strong liquid 
with $D$ following the Arrhenius law, $D=D_0\exp{(-E_A/k_BT)}$.

Along paths $\delta$, at the critical pressure the diffusivity jumps
from high-temperature fragile behavior to low-temperature strong
behavior close to the estimated temperature of the LLCP. An analogous
jump is found when crossing the LDL LMS line along paths
$\epsilon$. We note that the jump appears only when passing the LLCP or a
LMS line.  In analogy with the hysteresis behavior observed for the
isobaric specific heat (see Fig.~\ref{fig:2}), a jump is also observed
when heating up the systems along paths $\epsilon '$, but now it occurs
at a temperature corresponding to the HDL LMS line. This indicates that
the hysteresis caused by the first-order transition observed when
studying the thermodynamic behavior in a finite system is observed for
the diffusive behavior as well. This jump in the diffusion coefficient 
is a finding 
that can be useful for the experimental determination of
the liquid-liquid coexistence.
These findings confirm the coupling between the dynamic and the thermodynamic behavior
observed in bulk water ($P>P_c$ corresponds to $P<P_c$ and vice versa 
in other models~\cite{poole11} for the reversed slope of the coexistence line) and extends its
validity to hydrophobic solutions.

\begin{figure}[htbp]
\centerline{\psfig{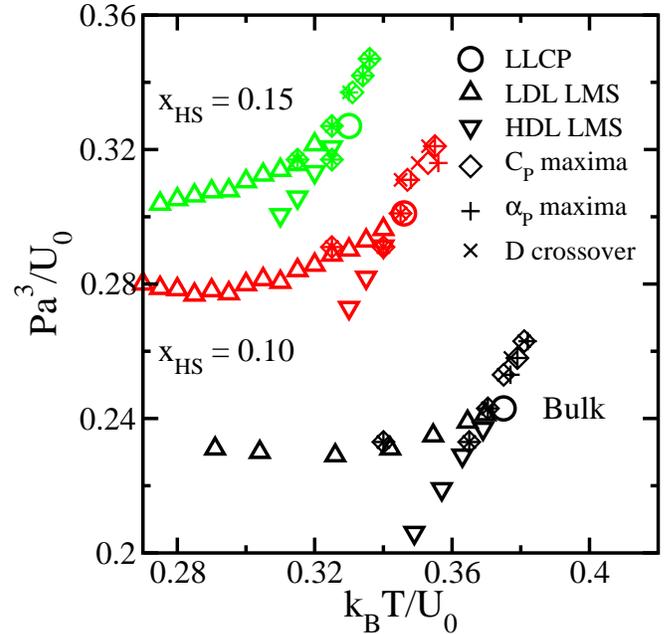}}
\caption{(Color online) Liquid-liquid phase diagram for bulk Jagla particles,
  $x_{\rm HS}=0.10$ and $x_{\rm HS}=0.15$ HS(J) solutions.  Along with the LLCP
  (bold large circles), LDL LMS (triangles up) and HDL LMS (triangles
  down), are shown the maxima of the isobaric specific heat $C_P$ (small
  circles), the maxima of the coefficient of thermal expansion
  $\alpha_P$ ($+$) and the crossover points of the diffusion coefficient
  $D$ ($\times$).  }
\label{fig:4}
\end{figure}

Figure~\ref{fig:4} shows the LLCP, HDL, and LDL LMS lines, the points at
which the isobaric specific heat $C_P$ and coefficient of thermal
expansion $\alpha_P$ maxima occur and the points at which a crossover in
the diffusive behavior is found. The coefficient of thermal expansion
$\alpha_P= (1/V)\cdot (\partial V/\partial T)_P$ has been calculated for
all our systems and for all paths. It displays a behavior qualitatively
equivalent to that of the isobaric specific heat. Above the
LLCP along the Widom line $C_P$ and $\alpha_P$ exhibit maxima, and the
temperature at which the diffusivity crossover occurs, estimated from
the derivative of the logarithm of $D$ with respect to $1/T$, coincides
with $C_P$ and $\alpha_P$ temperature maxima and thus with the Widom
line. The maxima in thermodynamic response functions and the diffusivity
crossover at the critical pressure are also found close to the LLCP
temperature, estimated in $NVT$ simulations \cite{corradini10pre,
  xu2}. Below the LLCP and along paths $\epsilon$ and $\epsilon '$ the
points where $C_P$ and $\alpha_P$ and the diffusivity crossover occurs
also coincide, and they correspond to the crossing of the LDL LMS line
and the HDL LMS line, respectively. This shows that the hysteresis
expected for the thermodynamic quantities calculated along paths
$\epsilon$ and $\epsilon '$ due to the first-order coexistence line
between LDL and HDL and observed only at the LMS lines due to the
metastability of the phases, is also maintained in the diffusive
behavior. 

\section{Conclusions}\label{conc}

Our results suggest that the FTS dynamic
transition reported for water in different environments in the
literature \cite{xu,gotze2010,sciortinogallo96,block7,WL1,longinotti}
occurs at the same time as the thermodynamic liquid-liquid transition.
They are, in other words, two sides of the same coin and, when the
isobaric paths do not cross the LMS or the Widom line, there is no
crossover and the dynamic behavior remains strong or fragile.  We note that
this connection between dynamic and thermodynamic behaviors might be a
general feature of network-forming liquids, e.g., a similar picture is
found in silica \cite{naturefrancesco}.  Why the FTS
dynamic crossover is also commonly found in glass formers~\cite{block8} when it is not
associated to a liquid-liquid transition remains an open question.
Our results show that for moderately concentrated solutions the changes induced by the solutes
are continuous and the FTS crossover remains coupled to the liquid-liquid transition 
paving the way to the experimental exploration
of aqueous solutions as tools to understand the mysteries of water.
In particular light scattering and neutron scattering experiments~\cite{liu} 
can be used to detect the liquid-liquid transition in
aqueous mixtures. Particularly important would be to see the jump in relaxation
times below the LLCP as it is
directly connected to the liquid-liquid transition. 

\medskip

\section*{Acknowledgements}
D.C. and H.E.S. thanks the NSF Chemistry Division for support (grants CHE-0404673, CHE
0911389 and CHE 0908218).
D.C. and P.G. gratefully acknowledge the computational
support received by the INFN RM3-GRID at Roma Tre University. S.V.B.
acknowledges the partial support of this research through the Dr.
Bernard W. Gamson Computational Science Center at Yeshiva College. 


\end{document}